\documentclass[12pt]{article}
\usepackage[english]{babel}
\usepackage{amsmath,amsthm,amssymb,amsfonts,latexsym}
\usepackage{color}

\usepackage{graphicx}
\usepackage{subfigure}
\catcode `\@=11 \@addtoreset{equation}{section}

\catcode `\@=12

\begin{document}

\title{$(H,\rho)$--induced political dynamics: 
facets of the disloyal attitudes 
into the public opinion}  

\author{R.~D.~Salvo, M.~Gorgone and F.~Oliveri\\
\ \\
{\footnotesize Department of Mathematical and Computer Sciences,}\\
{\footnotesize Physical Sciences and Earth Sciences, University of Messina}\\
{\footnotesize Viale F. Stagno d'Alcontres 31, 98166 Messina, Italy}\\
{\footnotesize rdisalvo@unime.it; mgorgone@unime.it; foliveri@unime.it}
}

\date{Published in \textit{Int. J. Theor. Phys.} \textbf{56}, 3912--3922 (2017).}

\maketitle

\begin{abstract}                                                                                                                                                                                                                                                        
A simple model, suitable to describe the dynamics of a political system consisting of three macro--groups affected by turncoat--like behaviors and the influence of the opportunistic attitudes of
politicians on voters' opinion, is presented.
The model is based on raising and lowering fermionic operators whose dynamics is ruled by a suitable quadratic Hamiltonian operator with the addition of specific rules (depending on the variations of the mean values of the observables) able to adjust periodically the model to the political environment, \emph{i.e.}, we move in the framework of the so called
$(H,\rho)$--induced dynamics approach.
\end{abstract}


\section{Introduction}
\label{sect1}

In recent years, the mathematical description of the dynamical aspects of political systems, \emph{e.g.}, the creation of coalitions or cooperations, political decision making, voting 
rules and vote maximizers, attracted the attention of many researchers (see \cite{Johnson_Book} for 
a discussion about various mathematical approaches used in political science). Also, either classical epidemiological approaches or typical game theory tools have been used to build mathematical models for the analysis of spread of political parties, as well as of coalition formation and  change of voters' opinions over time 
(see \cite{Lichtenegger,Misra,Sened}, and references therein). 
Moreover, quantum--like models, \emph{i.e.}, models based on the mathematical formalism of quantum mechanics and its generalizations, recently proved successful in the fields of cognitive and social sciences, psychology, economy, finances, and game theory \cite{KhrennikovBook2004,KhrennikovBook2010,HavenKhrennikov2009,BusemeyerBook2012}. In particular, the methods of quantum information theory, as well as the formalism of the theory of open quantum systems, have been used to describe the dynamics of the voters' mental states and the approaching of a stable state of a decision equilibrium \cite{Khrennikova_Haven,Khrennikova}. 

In this paper, we consider a simplified model of a political system whose parties are characterized by the tendency of part of their members (called \emph{turncoats}) to change, often in an unscrupulous way
and repeatedly, allegiance, and how
this behavior may influence the voters' opinions. The main ingredients of the model are fermionic operators 
\cite{Merzbacher,Roman}, and the dynamical aspects, ruled by a self--adjoint time--independent 
Hamiltonian, are faced by using the Heisenberg representation enriched by using some rules acting 
periodically on the system on the basis of its state variations, so that we are in the
framework of $(H,\rho)$--induced dynamics \cite{DiSalvoOliveriAAPP2016,BDSGO_QGOL2017,BDSGO_Hrho2016}.
  
The description of the dynamics of macroscopic complex systems  through the use of \emph{raising} and \emph{lowering} operators and the number representation \cite{BagarelloBook} has been used 
in several recent papers to analyze the dynamical aspects in rather different areas
\cite{DiSalvoOliveriAAPP2016,BagarelloBook,BagarelloOliveriMigration2013,BagarelloHaven2014a,BagarelloHaven2014b,BagarelloOliveriEco2014,BagarelloGarganoOliveriEscape2015,BagarelloCherubiniOliveriDesert2016,DiSalvoOliveriRM2016}, and also to describe alliances in politics \cite{BagarelloAlliance2015,BagarelloHavenAlliance2016} with  reference to the
Italian political system \cite{DSO_turncoat,DSGO_turncoat}. 

By observing the contemporary political landscape in place in Italy, a look at the behavior of 
politicians from various 
parties during the first thirty months of the XVII Italian Legislature (started in 2013) reveals a high level of disloyal attitude and openness towards accepting chameleons
from other political groups in both the two houses of the Italian Parliament; from the official data available in the institutional web pages of the Chamber of Deputies 
({http://www.camera.it}) and the Senate of the Republic ({http://www.senato.it}), 
more than 300 jumps between different parties during the period of 30 months has been recorded; in \cite{DSO_turncoat}, it has been shown that  the numerical results of the operatorial model there considered gave a 
satisfactory fit with these data. The general political elections of 2013 in Italy produced a result where three parties (the \emph{Partito Democratico}, 
PD, the \emph{Popolo della Libert\`a}, PdL, and the \emph{Movimento 5 Stelle}, M5S) took more or less the same 
number of votes, while other political groups gained few seats in the Parliament. PD was really the first party in that 
election, but due to the different electoral rules for the two houses of Parliament, it got the majority of parliamentary seats only in the Chamber of Deputies. Then, the 
formation of a stable government required the birth of alliances between different parties. 
This determined a fragmentation of some political parties, the rise of new parliamentary groups, and a paroxysmal occurrence of exchanges of of parliament members among different political groups.  

Our model consists of nine groups of politicians: once three possible simplified political strategies have been identified (representing a moderate, rather than a fickle or an extremist attitude, respectively), each main group is thought of as composed by three subgroups; these latter subgroups include the politicians who are loyal to their own group, and the ones, belonging to the same party, which, instead, are drawn towards another of the two remaining groups.
Moreover, the description of voters' opinions is modeled by considering the supporters of the parties, as well as an additional compartment for the abstainers, certified by all recent polls to be the largest group of voters.   

Roughly speaking, our model is made by two compartmental submodels, each described by fermionic operators and whose evolution is driven by a time--independent self--adjoint Hamiltonian operator. These two submodels are, in a certain sense, coupled by means of a rule suitable to adjust the model according to the current state reached by the system. This method ($(H,\rho)$--induced dynamics, \cite{DiSalvoOliveriAAPP2016,BDSGO_QGOL2017,BDSGO_Hrho2016}), even if the system under consideration is closed, is able to produce dynamical behaviors admitting asymptotic states. In general, this is not the unique way to proceed in order to get non--oscillating evolutions; for instance, in \cite{Khrennikova}, along the lines developed in \cite{Asano2011,Asano2012,AsanoBook2015}, the open quantum system theory has been adopted to model the process of establishing the so called political behavioral equilibrium by using the Markov approximation of the quantum master equation. Contrarily to what happens when one considers open quantum systems, our approach deals with finite--dimensional Hilbert spaces, and the definition of the rule allows us to introduce in the model phenomenological aspects which are not easy to include in any hermitian time--independent Hamiltonian operator. It is worth to be noticed that the approach described in this paper in some sense provides results similar to those obtained by deriving the dynamics by means of a time--dependent Hamiltonian operator, \cite{BDSGO_Hrho2016}.

The plan of the paper is the following one. 
In Section~\ref{sect2}, we briefly sketch the operatorial framework and the main ingredients of the so called
$(H,\rho)$--induced dynamics \cite{DiSalvoOliveriAAPP2016,BDSGO_QGOL2017,BDSGO_Hrho2016}; then, we introduce our model. 
Finally, Section~\ref{sect3} contains some numerical simulations and a brief discussion of the results.

\section{The mathematical framework and the model}
\label{sect2}
Consider an $N$--compartment model, and associate  to each compartment an annihilation ($a_j$), a creation ($a^\dagger_j$), and a number ($\hat n_j=a^\dagger_j a_j$) operator. These actors satisfy the \emph{Canonical Anticommutation Rules} \cite{Merzbacher,Roman}
\begin{equation}
\{a_{j},a^\dagger_{k}\}=\delta_{j,k}\mathbb{I},\quad 
\{a_{j}, a_{k}\}=\{a^\dagger_{j}, a^\dagger_{k}\}=0, \qquad j,k=1,\ldots,N,
\end{equation}
where $\{x,y\}=xy+yx$.
The states of the system are vectors in the $2^N$--dimensional Hilbert space 
$\mathbb{H}$ constructed as the linear span of the vectors
\begin{equation}
\phi_{n_1,n_2,\ldots,n_N}:=(a_1^\dagger)^{n_1}(a_2^\dagger)^{n_2}\cdots(a_N^\dagger)^{n_N}\phi_{\bf 0},
\end{equation}
where $n_j \in \{0,1\}$ for all $j=1,\ldots,N$, and $\phi_{\textbf{0}}$ is the \emph{vacuum} of the theory, \emph{i.e.},  a vector annihilated by all the operators $a_j$.
The vectors $\phi_{n_1,\ldots,n_N}$ give an orthonormal set of eigenstates of the number operators,
say
\begin{equation}
\hat n_j \phi_{n_1,\ldots,n_N}=n_j \phi_{n_1,\ldots,n_N}, \quad \mbox{for all } j=1,\ldots,N.
\end{equation}

Let $H=H^\dagger$ be a time--independent Hamiltonian operator embedding the main effects deriving from the interactions among the compartments.
The dynamics of any operator $X$ is deduced in the Heisenberg representation,
$X(t)=\exp(\mathrm{i}Ht)X\exp(-\mathrm{i}Ht)$; therefore,  the dynamics is obtained by computing, at each instant $t$, 
the exponential of a $2^N\times 2^N$ matrix. The same task can be achieved
by solving a system of $N 4^N$ differential equations, say
\begin{equation}
\dot a_j(t)=\mathrm{i}[H,a_j(t)], \qquad j=1,\ldots,N,
\end{equation}
where $[\cdot,\cdot]$ is the commutator, and each $a_j$ is a $2^N\times 2^N$ matrix: for large $N$ this could be a serious problem.
The computational cost reduces drastically if the Hamiltonian is quadratic, \emph{e.g.},
\begin{equation}
H=\sum_{j=1}^N\omega_j a_j^\dagger a_j+\sum_{1\le j< k\le N}\lambda_{jk}(a_j^\dagger a_k+a_k^\dagger a_j).
\end{equation}
In fact, by introducing 
\begin{equation}
A(t)=\left(
\begin{array}{c}
a_1(t)\\
a_2(t)\\
\ldots\\
a_N(t)
\end{array}
\right), \qquad
U=\mathrm{i}\left(
\begin{array}{ccccc}
-\omega_1 & \lambda_{12} & \lambda_{13} & \cdots & \lambda_{1N}\\
\lambda_{12} & -\omega_2 &  \lambda_{23} & \cdots & \lambda_{2N}\\
\cdots & \cdots & \cdots & \cdots & \cdots \\
\lambda_{1N} &\lambda_{2N} &\lambda_{3N} & \cdots & -\omega_N
\end{array}
\right),
\end{equation}
we may write
$\dot A(t)=U A(t)$,
whereupon
$A(t)=V(t)A(0)=\exp(Ut)A(0)$.

Once defined a vector state 
$\phi_{n_1,\ldots,n_N}$ representing the initial configuration of the system, we compute the mean values
\begin{equation}
\label{time_evolution}
n_j(t)=\langle\phi_{n_1,\ldots,n_N},\,
\hat n_j(t)\phi_{n_1,\ldots,n_N}\rangle, \qquad j=1,\ldots,N,
\end{equation}
$\langle\cdot,\cdot\rangle$ being  the scalar product in $\mathbb{H}$.

These average values can be  interpreted as a measure of the densities of the compartments of the model
\cite{BagarelloOliveriMigration2013,BagarelloOliveriEco2014}.
If $n_\ell$ is the initial density of the $\ell$--th compartment, we have 
\begin{equation}
n_j(t) = \sum_{\ell=1}^{N}\left|V_{j\ell}(t)\right|^2\,n_\ell,
\end{equation}
\emph{i.e.}, the local densities at time $t$ of the various compartments in a simple manner at the cost of computing the exponential of the $N\times N$ matrix $Ut$. 

It is easily ascertained that, by using a quadratic Hamiltonian, 
the dynamics we can deduce is at most quasiperiodic.
We can enrich the description by introducing some rules 
repeatedly acting on the parameters involved in the Hamiltonian
to account for a sort 
of dependence of the parameters on the current state of the system ($(H,\rho)$--induced dynamics
\cite{DiSalvoOliveriAAPP2016,BDSGO_QGOL2017,BDSGO_Hrho2016}). 

Consider a time interval $[0,T]$, and split it in $n=T/\tau$ subintervals of length $\tau$.
In the $k$--th subinterval $[(k-1)\tau,k\tau[$, $k=0,\ldots,n$,  take a Hamiltonian $H^{(k)}$
ruling the dynamics. The global dynamics arises from a sequence of Hamiltonians,
\begin{equation}
H^{(1)} \stackrel{\tau}{\longrightarrow} H^{(2)} \stackrel{\tau}{\longrightarrow} H^{(3)} \stackrel{\tau}{\longrightarrow} \cdots \stackrel{\tau}{\longrightarrow} H^{(n)},
\end{equation}
all having the same functional form but possibly involving different values of some of their parameters.
The latters are changed according to the variations of the actual state of the system in a subinterval; thus,  
the values of the inertia parameters of some compartments are repeatedly adjusted, so that asymptotic behaviors for the evolution of the mean values can be recovered \cite{BDSGO_Hrho2016}. 

Consider  a self--adjoint quadratic Hamiltonian operator $H^{(1)}$, the evolution of an observable $X$,
$X(t)=\exp(\mathrm{i}H^{(1)}t)X\exp(-\mathrm{i}H^{(1)}t)$,  
and its mean value
$x(t)=<\phi_{n_1,\ldots, n_N},X(t)\phi_{n_1,\ldots, n_N}>$
in a time interval of length $\tau>0$.
Change some of the parameters involved in $H^{(1)}$ on the basis of the values of the various $x(\tau)$, thus getting  a new Hamiltonian operator $H^{(2)}$, having the same functional form as $H^{(1)}$.
Continue the evolution of the system as ruled by $H^{(2)}$ for one more time interval of length 
$\tau>0$, and so on.
In every subinterval we have a system like
\begin{equation}
\label{dynamics}
\dot A(t)=U^{(k)}A(t),  \qquad t \in [(k-1)\tau,k\tau[.
\end{equation}
Since what we really need is $\exp(U^{(k)}t)$, let us consider in \eqref{dynamics} $A(t)$ as an $N\times N$
matrix such that $A(0)=\mathbb{I}$, whence the solution of \eqref{dynamics} is just 
\begin{equation}
A(t)=\exp(U^{(k)}t),\qquad t \in [(k-1)\tau,k\tau[,
\end{equation}
whereupon, gluing the solutions in the various subintervals, we have globally
\begin{equation}
A(t)=\left\{
\begin{array}{lll}
\exp(U^{(1)}t) &\qquad & t \in [0,\tau[\\
\exp(U^{(2)}(t-\tau))\exp(U^{(1)}\tau) & & t \in [\tau,2\tau[\\
\exp(U^{(3)}(t-2\tau))\exp(U^{(2)}\tau)\exp(U^{(1)}\tau) & & t \in [2\tau,3\tau[\\
\cdots & & \cdots
\end{array}
\right.,
\end{equation}
and so we can compute the complete evolution of the mean values.

\subsection{The model of the political system and of voters' opinions}
\label{subsect:opinions}
\begin{figure}[h]
\centering
\includegraphics[width=0.9\textwidth]{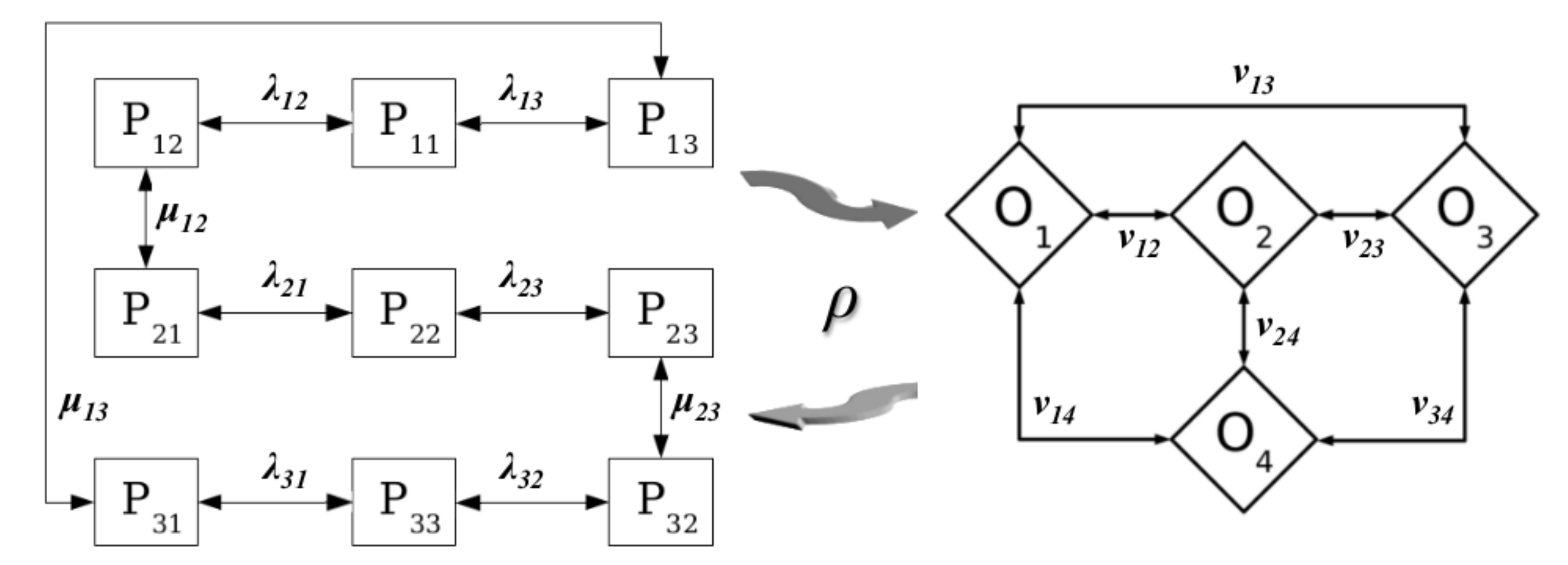}
\caption{\label{fig:model2} A schematic view to the general context.}
\end{figure}

The model  (schematized in Fig.~\ref{fig:model2}) describes a system made of two
subsystems. The first one consists of nine groups of politicians $P_{ij}$ ($i,j=1,2,3$),
represented by fermionic operators, identified on the basis of three possible attitudes 
in relation to three main factions characterized by three possible simplified 
political strategies. More precisely, at the initial time, the first faction represents a moderate 
group, characterized by scarce openness to exchanges with other parties (external 
flows) and small propensity to transformations in ideological direction (internal flows), 
the second faction represents a fickle group, highly permissive towards transitions between 
political groups and other ideology sympathizers in the party itself, while the third faction 
stands for an extremist group, intransigent against contaminations and influences 
from others. The operators used to describe the evolution of the system are to be 
intended in accordance with the key under which the compartments whose subscript 
indices are equal refer to those politicians who are loyal to the corresponding faction, 
while those having different subscript indices are related to the politicians who belong 
to a faction (the one associated with the first index) but are drawn towards another 
faction (the one associated with the second index). In the second subsystem, the actors $O_j$ ($j=1,\ldots,4$) are again fermionic operators, whose mean values refer to the amounts of the $j$--th party 
supporters for $j=1,2,3$, and of the indecisive people choosing not to vote for $j=4$, respectively.
The time evolution of the densities described by the number operators $O^\dagger_j O_j$ ($j=1,\ldots,4$) thus offers a view on the trends of support (or dissatisfaction) of the various voters shown at the polls.

The observables we are interested to are the following:
\begin{equation}
\label{eq:observables}
\begin{aligned}
&f_i(t)=\sum_{j=1}^3p_{ij}(t)=\sum_{j=1}^3P_{ij}^\dagger(t)P_{ij}(t), \qquad i=1,\ldots,3,\\ 
&o_\ell(t)=O_\ell^\dagger(t)O_\ell(t),\qquad \ell=1,\ldots,4,
\end{aligned}
\end{equation}
that we interpret as instantaneous measures of the densities of the three political factions, $f_i$ $(i=1,2,3)$, the corresponding supporters, $o_\ell$ $(\ell=1,2,3)$, and the group of the abstainers, $o_4$, respectively. During the dynamics the two quantities $\sum_{i=1}^3 f_i(t)$ and $\sum_{\ell=1}^4 o_\ell(t)$ are conserved, since they both commute with the Hamiltonian $H$. In addition, a reasonable estimate of the majority $\mathcal{M}$, formed by the union of the leading faction $f_\ell$ and its sympathizers belonging to the unfair fringes inside the other factions, say 
\begin{equation}
\mathcal{M} = f_\ell+\sum_{\substack{k=1\\k\neq \ell}}^3 p_{k\ell},
\end{equation}
and the corresponding government--opposition 
\begin{equation}
\mathcal{O}=\sum_{i=1}^3 f_i - \mathcal{M}
\end{equation}
are taken into account.

The dynamics of the whole system is assumed to be governed by a self--adjoint time--independent Hamiltonian operator $H$ embedding the main effects deriving from the interactions among the compartments of the system,
\begin{equation}
\label{eq:Ham}
\begin{aligned}
H &= H^p+H^o,\\
H^p &= \sum_{i,j=1}^{3}\,\omega^p_{ij}\, P_{ij}^\dagger\,P_{ij}+\sum_{\substack{i,j=1\\ i\neq j}}^3 \lambda_{ij}\left(P_{ii}P_{ij}^\dagger+P_{ij}P_{ii}^\dagger\right)\\
&+\sum_{1\le i<j\le3} \mu_{ij}\left(P_{ij}P_{ji}^\dagger+P_{ji}P_{ij}^\dagger\right),\\
H^o&=\sum_{i=1}^4 \omega^o_i O_i^\dagger O_i+\sum_{\substack{i,j=1\\i\neq j}}^4 \nu_{ij}
(O_i^\dagger O_j+O_j^\dagger O_i).
\end{aligned}
\end{equation}

The real constants $\omega^p_{ij}$ and $\omega^o_{i}$ are 
related to the tendency of each degree of freedom to stay constant in time; 
the larger their values,
the smaller the amplitudes of the oscillations of the related densities 
\cite{BagarelloBook}; in
some sense, they are a measure of the \emph{inertia} of the various compartments. Moreover, the
real parameters $\lambda_{ij}$ are used to describe the internal flows that occur within each faction, 
namely the minor ideological positions representing in some sense early warning of disloyalty, whereas the 
parameters $\mu_{ij}$ are related to the external flows stemming from the real jumps between different parties; finally, 
$\nu_{ij}$ are related to the interchanges between different groups of supporters or a group of supporters and the group of abstainers.
 
Such a Hamiltonian reflects the circumstance that our system is actually made by two non--interacting subsystems; however, the interaction between the two systems is guaranteed by the introduction of the \emph{rule}. In fact, at fixed intervals of time, some of the parameters of the Hamiltonian are suitably
changed according to the state variation of the system. In particular, the inertia parameters of the
various compartments of the two subsystems are \emph{adjusted} as a consequence of the dynamics (see Section~\ref{sect3}).

\section{Numerical simulations}
\label{sect3}

\begin{figure}
\centering
\subfigure[]{\label{sr1}\includegraphics[width=0.48\textwidth]{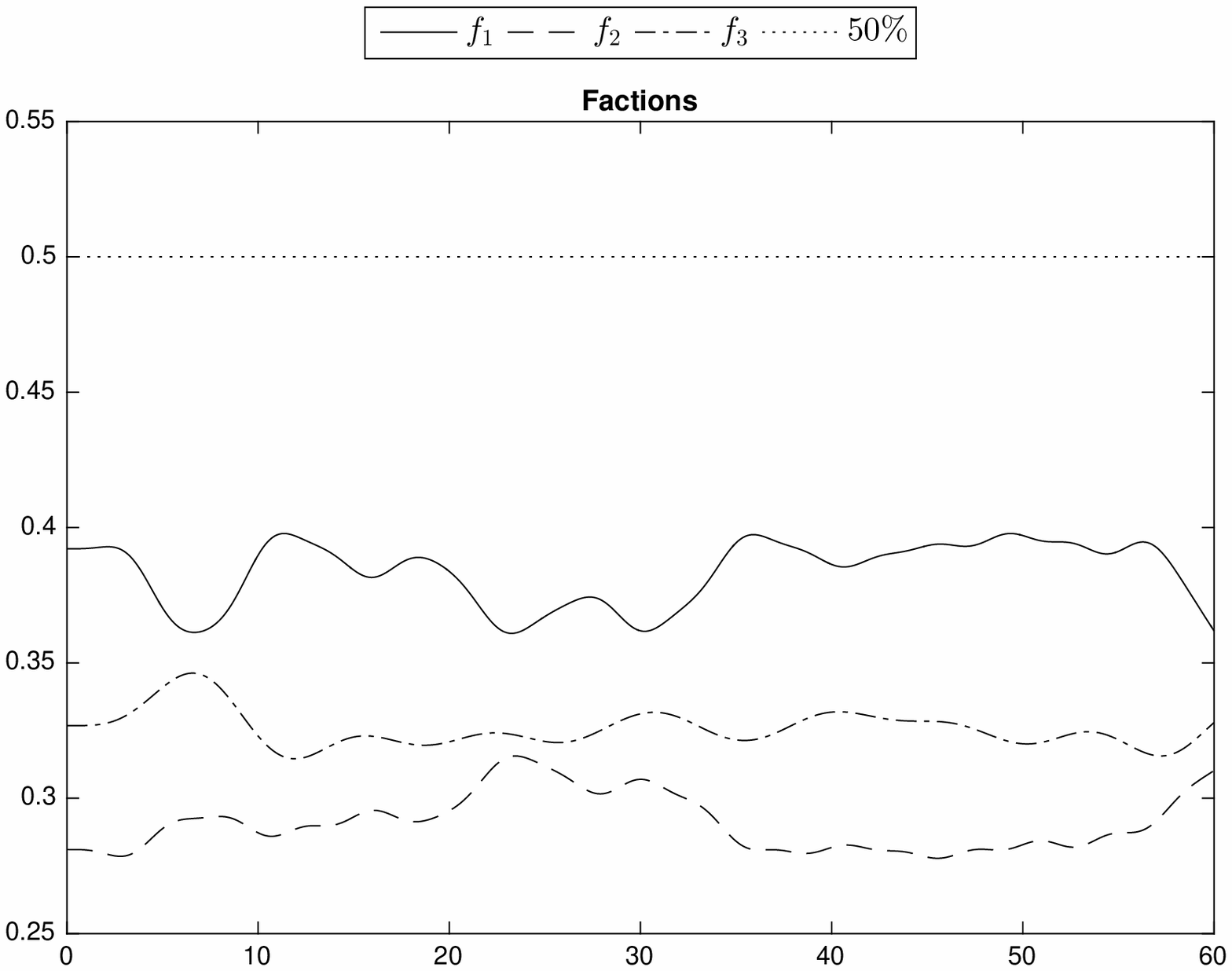}}\quad
\subfigure[]{\label{sr2}\includegraphics[width=0.48\textwidth]{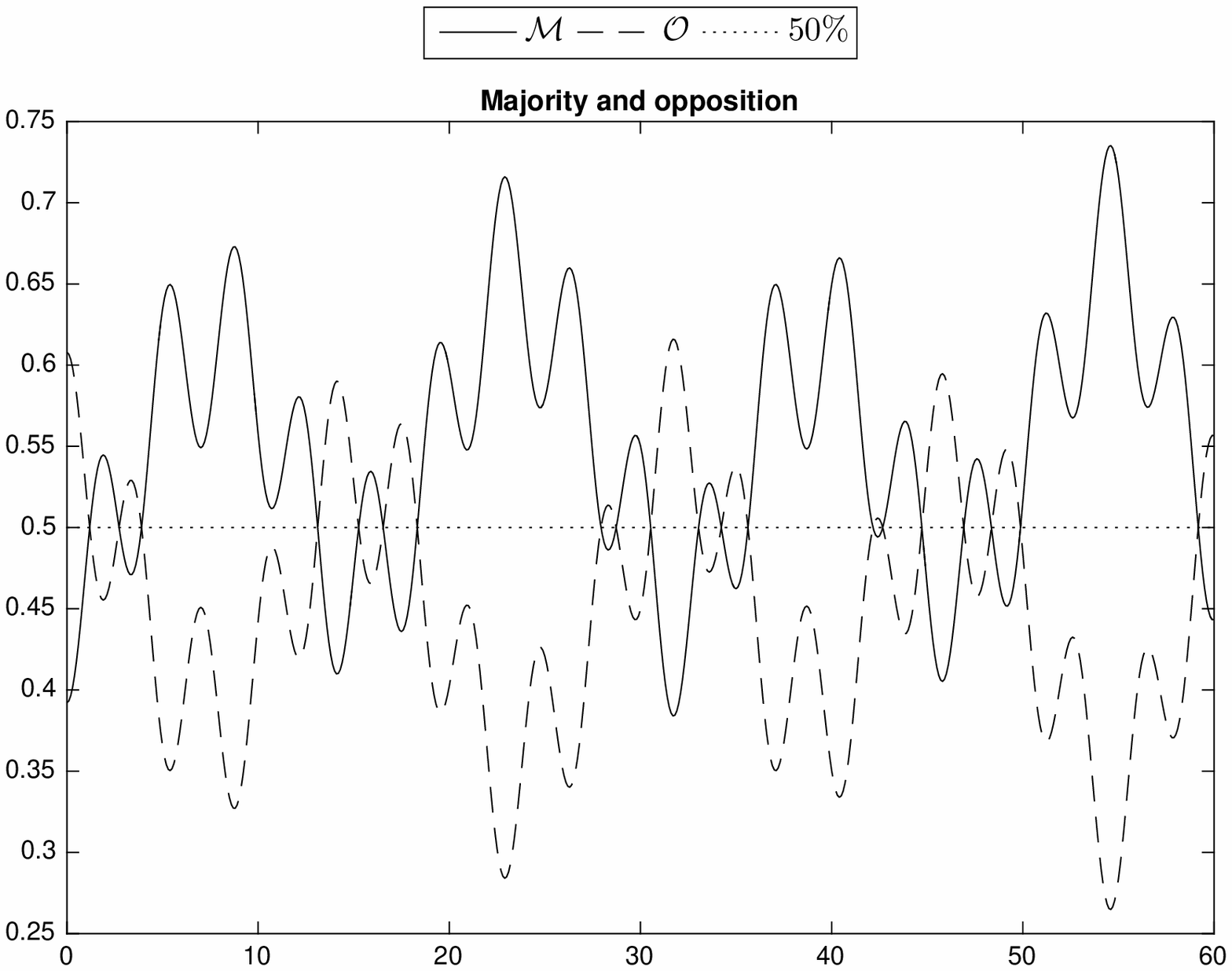}}\\\smallskip
\subfigure[]{\label{sr3}\includegraphics[width=0.7\textwidth]{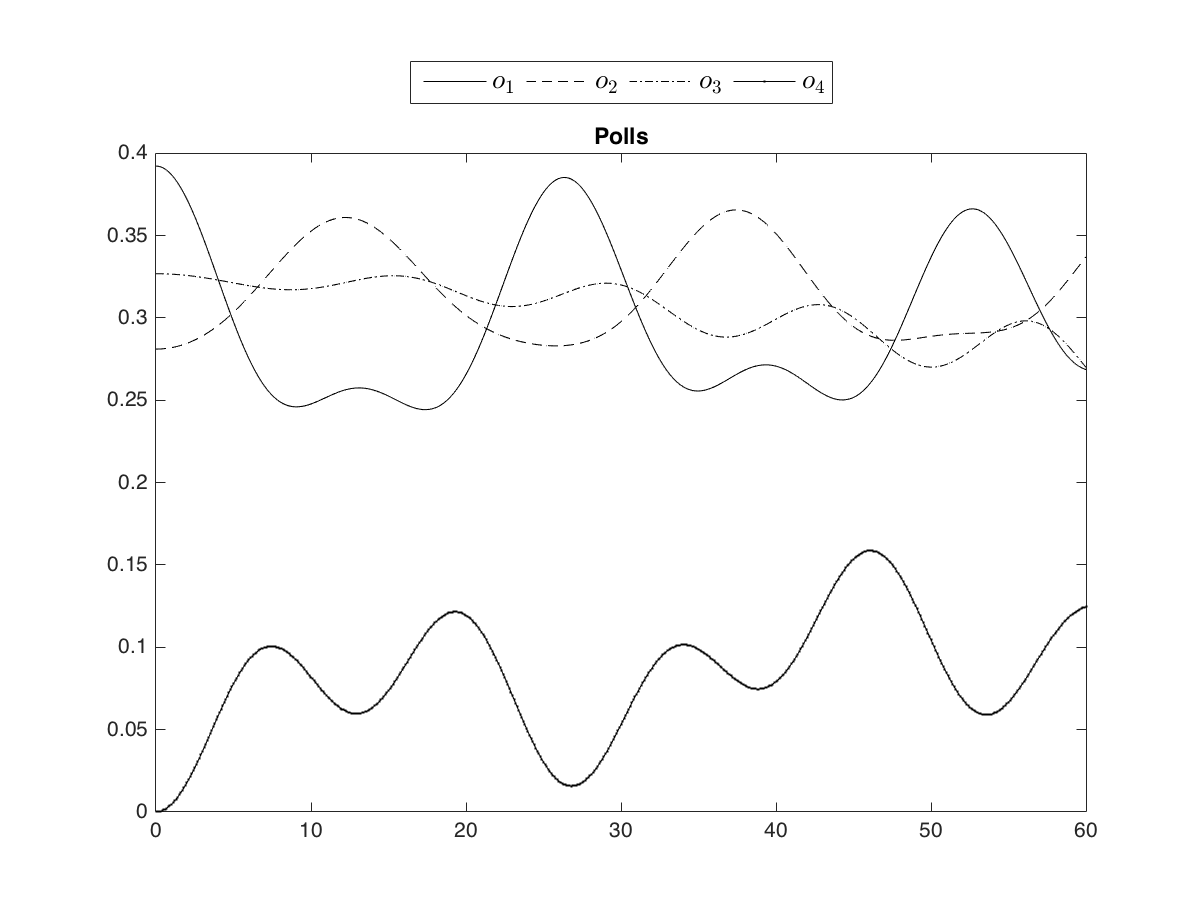}}
\caption{\footnotesize{\label{fig1} Linear model: dynamics of political parties affected by turncoat--like behaviors (frames \ref{sr1} and \ref{sr2}) and time evolution of the voters' opinions (frame \ref{sr3}) by using the initial condition \eqref{first:IC}. The polls show the support or dissatisfaction of the voters related to the behavior of politicians inside the government. The $x$--axis is scaled to the number of months in a five--year term of office. The plotted data are normalized to the sum of the initial densities.}}
\end{figure}

\begin{figure}
\centering
\subfigure[]{\label{r1_1}\includegraphics[width=0.48\textwidth]{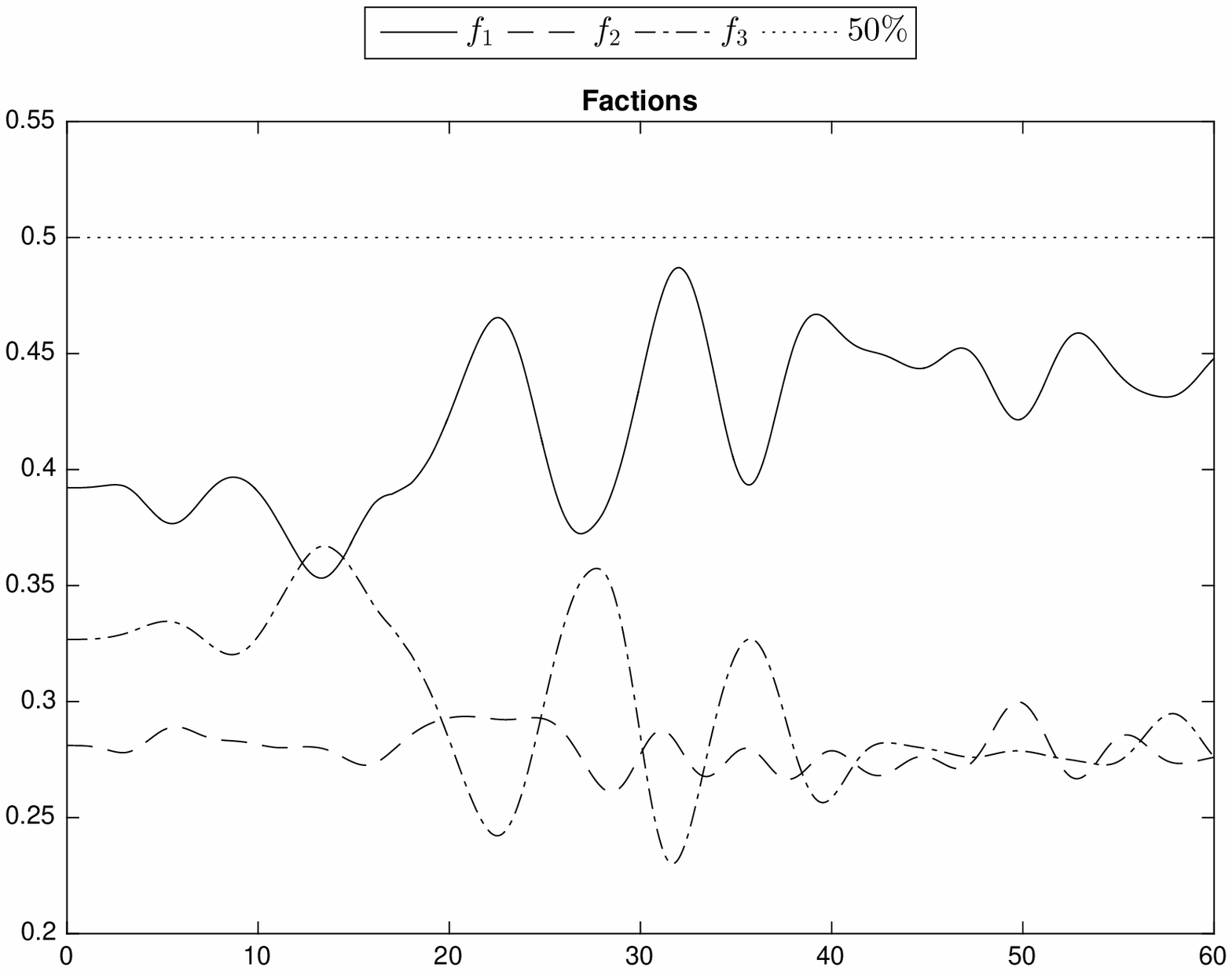}}\quad
\subfigure[]{\label{r2_1}\includegraphics[width=0.48\textwidth]{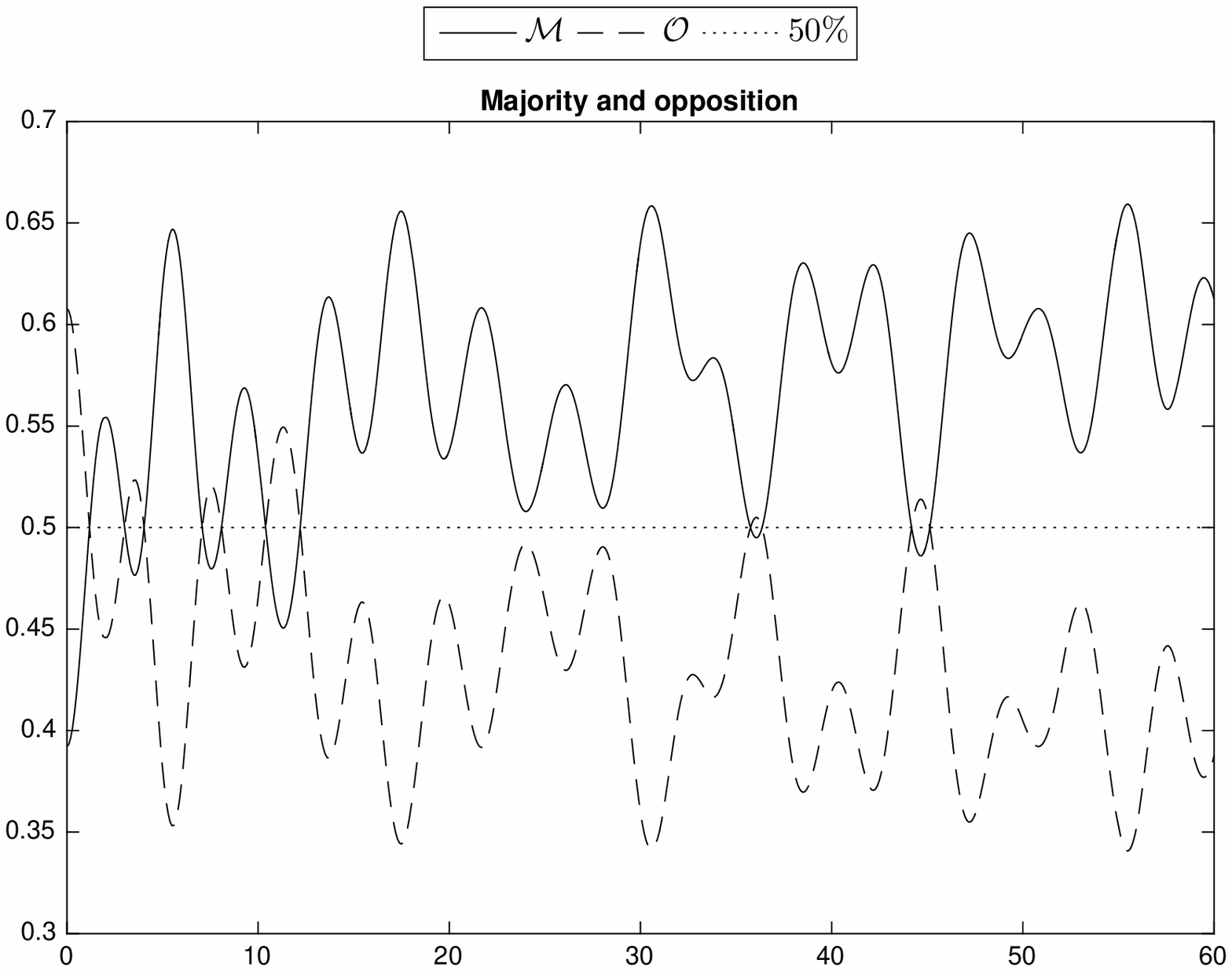}}\\\smallskip
\subfigure[]{\label{r3_1}\includegraphics[width=0.7\textwidth]{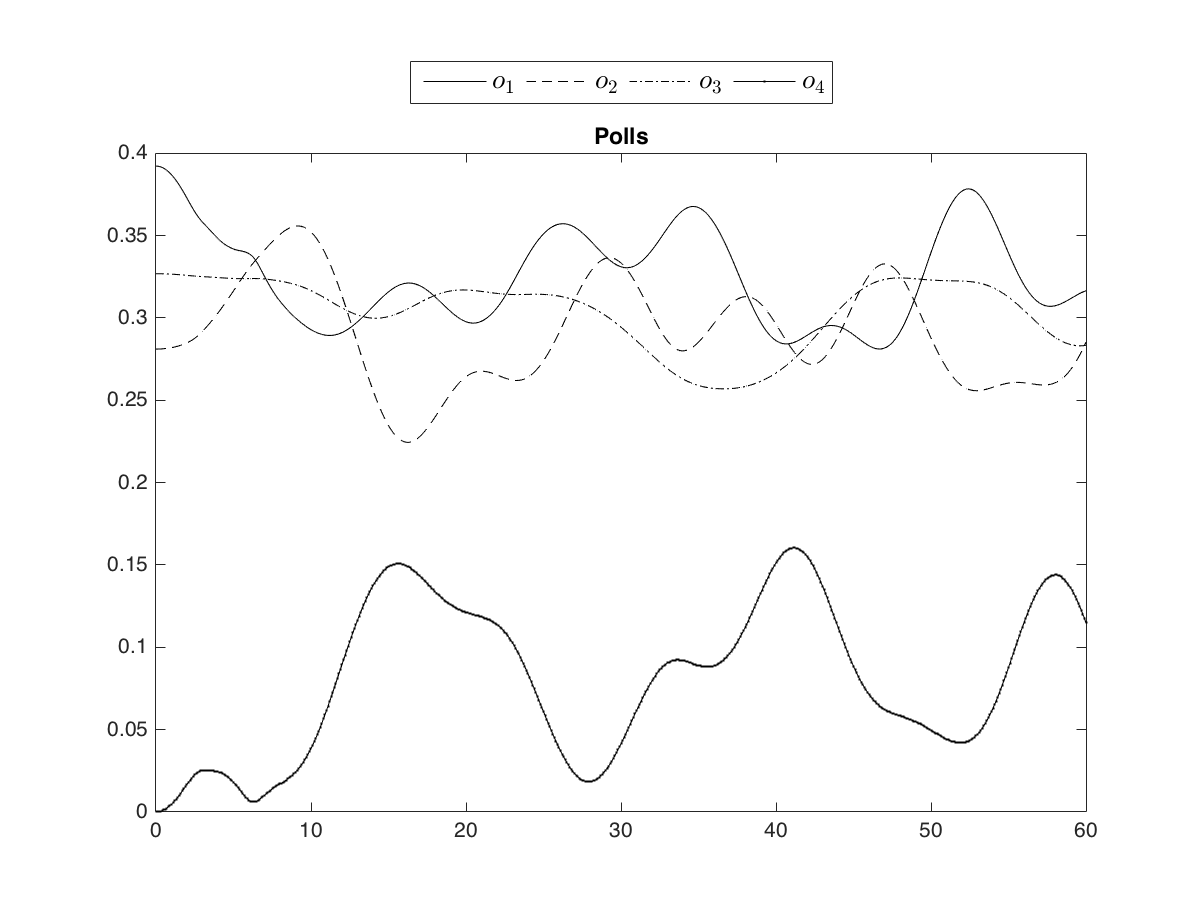}}
\caption{\footnotesize{\label{fig2} Stepwise linear model using the rule $\rho$ after each time step of length $\tau=1$: dynamics of political parties affected by turncoat--like behaviors (frames \ref{r1_1} and \ref{r2_1}) and time evolution of the voters' opinions (frame \ref{r3_1}) by using the initial condition \eqref{first:IC}. The polls show the support or dissatisfaction of the voters related to the behavior of politicians inside the government. The $x$--axis is scaled to the number of months in a five--year term of office. The plotted data are normalized to the sum of the initial densities.}}
\end{figure}

\begin{figure}
\centering
\subfigure[]{\label{r1_2}\includegraphics[width=0.48\textwidth]{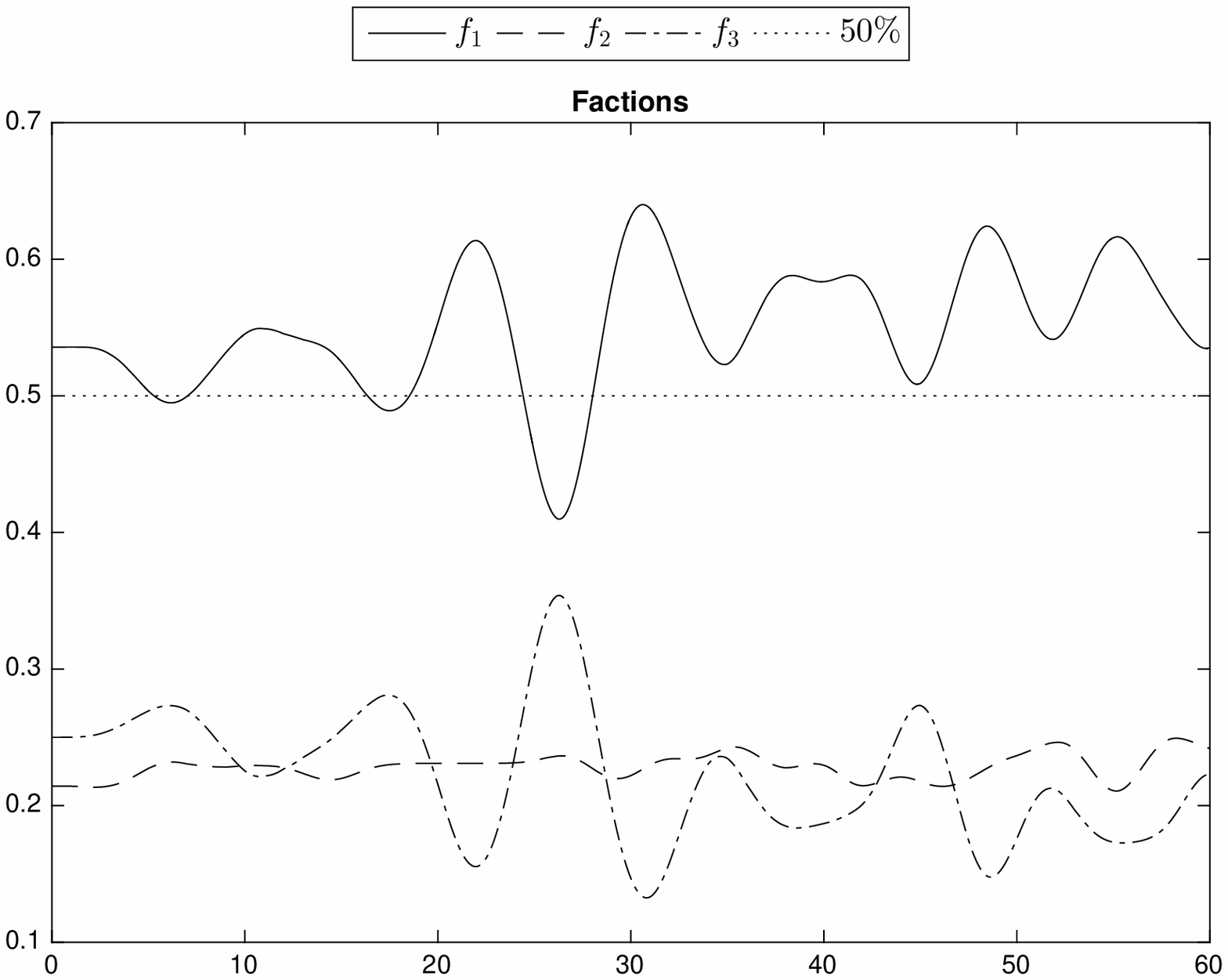}}\quad
\subfigure[]{\label{r2_2}\includegraphics[width=0.48\textwidth]{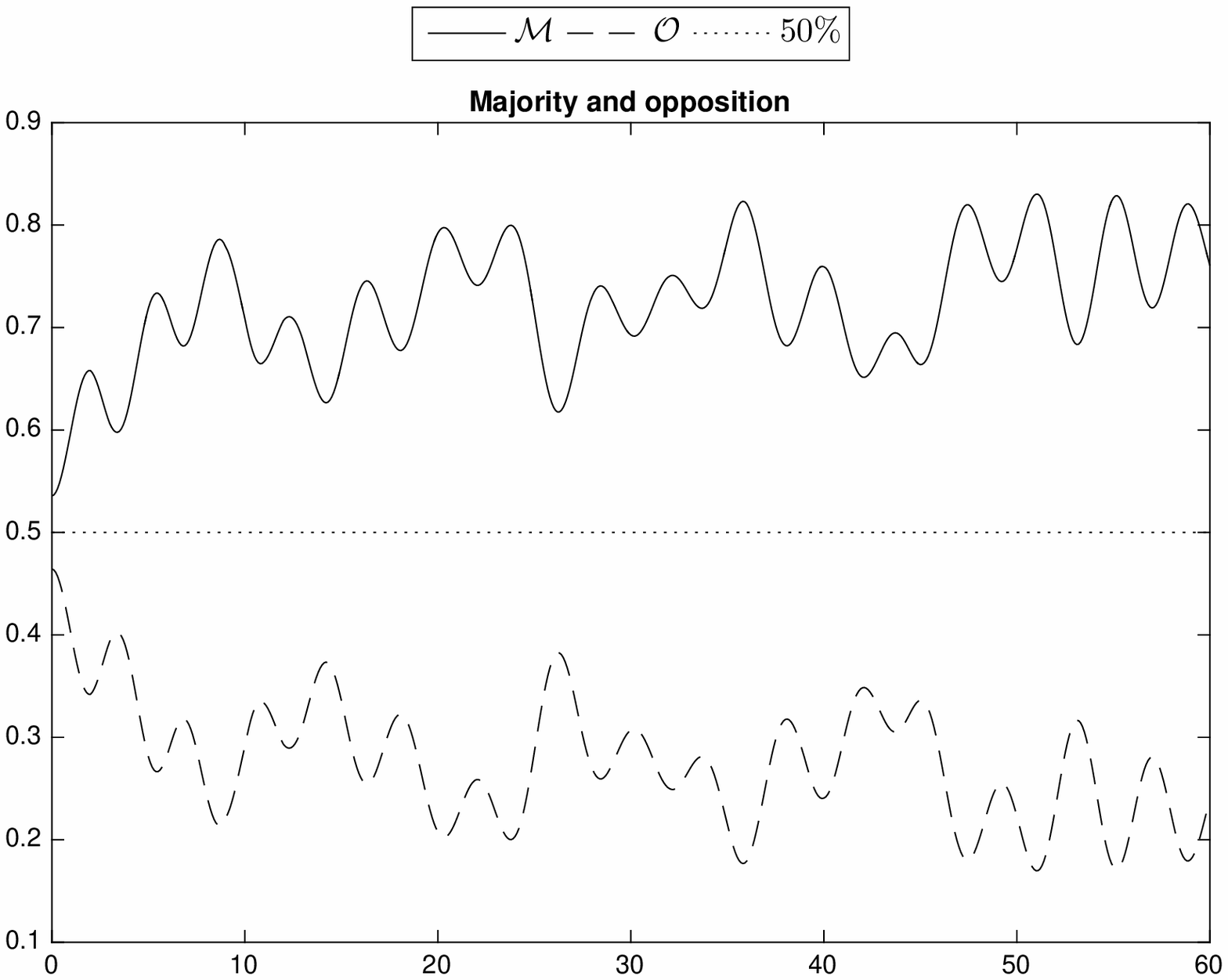}}\\\smallskip
\subfigure[]{\label{r3_2}\includegraphics[width=0.7\textwidth]{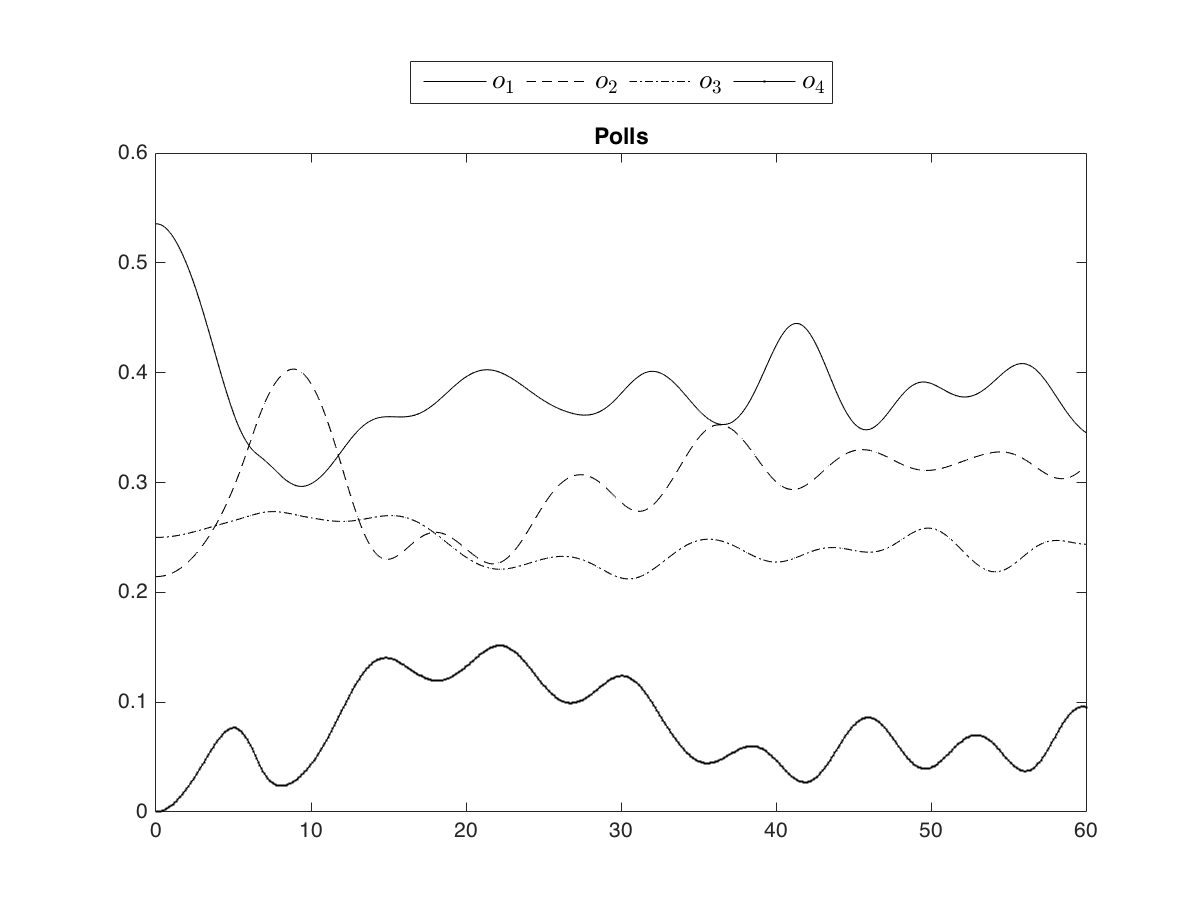}}
\caption{\footnotesize{\label{fig3} Stepwise linear model using the rule $\rho$ after each time step of length $\tau=1$: dynamics of political parties affected by turncoat--like behaviors (frames \ref{r1_2} and \ref{r2_2}) and time evolution of the voters' opinions (frame \ref{r3_2}). The initial condition \eqref{second:IC} of the trajectories of the densities of the three different factions of the system are such that the first faction possesses the majority of seats and the remaining two have similar small densities. The polls show the support or dissatisfaction of the voters related to the behavior of politicians inside the government. The $x$--axis is scaled to the number of months in a five--year term of office. The plotted data are normalized to the sum of the initial densities.}}
\end{figure}

According to the choice of the values of the parameters involved in the Hamiltonian, as well as to the definition of the \emph{rule},   several dynamical behaviors can be described. Of course, also the initial conditions play an important role. 

The values of the parameters used in the numerical simulations, chosen in such a way they 
mimic the different political styles of the three parties and the aptitude of their supporters, are the following ones:
\[
\begin{aligned}
&\omega^p_{11} = 0.7,\; 
\omega^p_{12} = 0.6,\; 
\omega^p_{13} = 0.5,\; 
\omega^p_{21} = 0.4,\; 
\omega^p_{22} = 0.5,\;
\omega^p_{23} = 0.3,\; 
\omega^p_{31} = 0.8,\\ 
&\omega^p_{32} = 0.9,\; 
\omega^p_{33} = 0.1,\;
\lambda_{12} = 0.25,\;
\lambda_{13} = 0.3,\;
\lambda_{21} = 0.6,\;
\lambda_{23} = 0.65,\\
&\lambda_{31} = 0.15,\;
\lambda_{32} = 0.1,\;
\mu_{12} = 0.1,\;
\mu_{13} = 0.09,\;
\mu_{23} = 0.01,\\
&\omega^o_{1} = 0.65,\;
\omega^o_{2}= 0.6,\;
\omega^o_{3} = 0.89,\;
\omega^o_{4} = 0.9,\;
\nu_{12} = 0.1,\; 
\nu_{13} = 0.07,\\ 
&\nu_{14} = 0.1,\;
\nu_{23} = 0.1,\; 
\nu_{24} = 0.05,\; 
\nu_{34} = 0.02.
\end{aligned}
\]

Two different initial conditions have been chosen:
\begin{eqnarray}
\label{first:IC}
p_{11}(0)=o_1(0)=0.6,\; p_{22}(0)=o_2(0)=0.43,\; p_{33}(0)=o_3(0)=0.5,\\ 
\label{second:IC}
p_{11}(0)=o_1(0)=0.75,\; p_{22}(0)=o_2(0)=0.3,\; p_{33}(0)=o_3(0)=0.35;
\end{eqnarray}
in both cases we assume initially that $p_{12}(0)=p_{13}(0)=p_{21}(0)=p_{23}(0)=p_{31}(0)=p_{32}(0)=o_4(0)=0$, meaning that at the beginning all members of a party are in the loyal subfaction.
The initial condition \eqref{first:IC} describes a situation where the three parties begin to interact with close initial densities, whereas the initial condition \eqref{second:IC} corresponds to a situation where a party has a density greater that the sum of the (close) densities of the remaining ones; moreover, the initial density of the supporters is the same as the one of the related party. 

The Heisenberg dynamics of the number operators in \eqref{eq:observables}, driven by the time--independent quadratic Hamiltonian operator \eqref{eq:Ham}, produces, as expected, the quasiperiodic trajectories shown in Fig.~\ref{fig1}. In order to depict more interesting and realistic situations, we extend the description of the dynamics by introducing a periodic check on the effects occurring during the time evolution, and by adapting the values of the parameters of the model to the subsequent states of the system according to an explicit prescription, \emph{i.e.}, a rule, in the $(H,\rho)$--dynamics approach \cite{BDSGO_Hrho2016}.

To be more precise, once the variations of the densities of the main factions
\begin{equation}
\delta_{i}=f_{i}(k\tau)-f_{i}((k-1)\tau),\quad i=1,\ldots,3,
\end{equation}
and of the densities of the supporters (abstainers for $\ell=4$)
\begin{equation}
\Delta_{\ell}=o_{\ell}(k\tau)-o_{\ell}((k-1)\tau),\quad \ell=1,\ldots,4
\end{equation}
have been computed in the $k$--th period of length $\tau$ of the Heisenberg--like evolution of the political system, the rule $\rho$ here considered consists in the replacements
\begin{equation}
\label{eq:rho}
\left\{
\begin{aligned}
&\omega^p_{ii}\stackrel{\rho}{\longrightarrow}\omega^p_{ii} (1+\delta_{i})(1-\mathrm{s}\, \Delta_4),\qquad
\omega^p_{ij}\stackrel{\rho}{\longrightarrow}\omega^p_{ij} (1-\mathrm{s}\, \Delta_j),\\
&\lambda_{ij}\stackrel{\rho}{\longrightarrow}\lambda_{ij} (1+\mathrm{s}\, \Delta_j),\\
&\omega^o_{i}\stackrel{\rho}{\longrightarrow}\omega^o_{i} (1+\delta_{i})(1+\mathrm{s}\, \Delta_i),\qquad
\omega^o_{4}\stackrel{\rho}{\longrightarrow}\omega^o_{4} (1+\mathrm{s}\, \Delta_4),
\end{aligned}
\right.
\end{equation} 
where $i,j=1,\ldots,3$, $j\neq i$ and 
$\mathrm{s}=\hbox{sign}(\delta_1 \delta_2 \delta_3)$.

Since the quantity $\sum_{i=1}^3 f_i(t)$ is conserved during the dynamics, it follows that the only possibilities for the evolution of the political system are the uncertain situation in which one faction decreases its density while the remaining groups grow ($\mathrm{s}<0$), or the case in which two factions decline while the remaining one rises ($\mathrm{s}>0$). The rule $\rho$ changes the values of the inertia of the hard core of each faction on the basis of both the variation of the faction itself and the trend of the number of abstainers, the inertia of the disloyal subgroups on the contrary of the internal interaction parameters according to the variations in the related public opinion, and has the effect of stabilizing the rising groups of supporters of growing factions. Similar sets of conditions acting on the parameters involved in the Hamiltonian have already been used to model complex dynamics of closed systems (see \cite{DiSalvoOliveriAAPP2016}, and \cite{BDSGO_Hrho2016} for a general discussion on the $(H,\rho)$--induced dynamics). Also in this context, the extension of the description of the dynamics of the model by means of the rule $\rho$ produces interesting results. As visible in Figs.~\ref{fig2} and \ref{fig3}, the rule allows to describe the stable
persistence of a government majority during all the term of office; on the contrary, the facets of the jumps between parties of unfair politicians on the public opinion reflect in the insurgence of a non negligible group of abstainers. 

The model considered in this paper, though very simple, is able to capture what is really observed in both houses of Italian Parliament during the current Legislature. In particular, a similar model discussed in \cite{DSO_turncoat}, suitably identifying some macro--groups, provided a good fit with the real data of movements of parliamentary members among different political groups. This phenomenon had strong social consequences on the voters, and a consistent group of people who declared to refuse to vote clearly emerged in all the polls.

By combining the results for the evolution of a political system spoiled by disloyalty and self--interest with the rule--induced dynamics \cite{BDSGO_Hrho2016} of the voters' opinion system, we get a dynamics for the polls of voters in which we can observe a general trend of surge and decline in support over time, with an alarming increase and consolidation of the number of electors who refuse to vote. 
The main aim of this paper was to build a simplified model for describing both the fragmentation of a minimal political system affected by turncoat's behavior, and the effects of these disloyal attitudes of politicians on the voters' opinions, including  the rise of a large group of abstainers.  Further investigations with more detailed models are planned in the near future.

\section*{Acknowledgments}
The research was partially funded by the Ph.D. School in Mathematics and Computer Science of the University of Catania.
The authors are grateful to the unknown referees whose comments and suggestions contributed to improve the quality of the paper.

\end{document}